\documentclass[useAMS,usenatbib]{mn2e}

\usepackage[utf8]{inputenc}
\usepackage{graphicx}
\usepackage{multicol}
\usepackage{color}
\usepackage{tabularx}
\usepackage{longtable}
\usepackage{graphics,epsfig,aas_macros,amsmath}
\newcommand{\software}[1]{\texttt{\small \MakeUppercase{#1}}}

\newcommand{\xmmn}{{\it XMM-Newton~\/}}

\newcommand{\rxte}{{\it RXTE~\/}}

\newcommand{\chandra}{{\it Chandra~\/}}
\newcommand{\suzaku}{{\it Suzaku~\/}}

\begin{document}

\title{Observation of variable pre-eclipse dips and disk winds in the eclipsing LMXB XTE J1710-281}

\author[G. Raman et al.]
{Gayathri Raman$^1$, Chandreyee Maitra$^{2,3}$ and Biswajit Paul$^1$ \\ 
$^1$Department of Astronomy and Astrophysics, Raman Research Institute, Sadashivanagar, Bangalore-560080, India\\
$^2$Laboratoire AIM, IRFU/Service d'Astrophysique - CEA/DSM - CNRS - Universite Paris Diderot, Bat. 709, \\
CEA-Saclay, 91191 Gif-sur-Yvette Cedex, France \\
$^3$Max-Planck-Institut f{\"ur} extraterrestrische Physik, Giessenbachstra{\ss}e, 85748 Garching, Germany}

\date{}

 \maketitle

 \begin{abstract}
We report the first detection of highly ionized Fe species in the X-ray spectrum of the eclipsing and dipping Low Mass X-ray 
Binary XTE J1710-281. Using archival \chandra and \suzaku observations, we have carried out a spectro-timing analysis of the source 
during three different epochs.
We compare the average orbital profile and obtain differences in pre-eclipse dip morphologies between different observation epochs.
We observe an orbit to orbit evolution of the dips for the first time in this source in both the \chandra observations, reflecting 
changes in the structure of the accretion disc in timescales of hours. We further perform intensity resolved spectroscopy 
for both the \chandra and the \suzaku data to characterize the changes in the spectral parameters from the persistent to the 
dipping intervals. We find that the absorbers responsible for the dips, can be best described using a partially ionized 
partial covering absorber, with an ionization parameter, log($\xi$) of $\sim$2. The photon index of the source remained at 
$\sim$2 during both the \chandra and the \suzaku observations. In the 0.6-9 keV \suzaku spectra, we detect a broad 0.72 keV Fe L-alpha 
emission line complex and two narrow absorption lines at $\sim$6.60 keV and $\sim$7.01 keV.
The highly ionized Fe line signatures, being an indicator of accretion disc-winds, has been observed for the first time in XTE J1710-281.

\end{abstract}


\section{Introduction}

Accreting Low Mass X-ray Binary (LMXB) systems contain a compact object in a binary configuration with a  
mass-donating low-mass companion star \citep{LC1980}. A small subset of the total LMXB population that have high inclinations (i$\sim$70-90$^\circ$)
are the dipping and eclipsing sources \citep{diaztrigo2006}. Out of the thirteen LMXB dipping sources (dippers from now) 
that host a neutron star, six systems show complete eclipses \citep{D'Ai}. These systems present a variety of spectro-temporal signatures 
that allow efficient means of probing the accretion disc, it's atmosphere and possible structures on it.\\

Absorbers responsible for producing the intensity dips are understood to be vertical structures present on the outer 
accretion disc, with asymmetric azimuthal distribution \citep{White_swank, Boirin}. Spectroscopic studies in a set of six dippers 
(EXO 0748-676, 4U 1254-690, 4U 1624-490, MXB 1659-298, 4U 1746-371 and 4U 1915-05) revealed that the spectral continuum during dips 
could be well described by simple ionized absorber models \citep{diaztrigo2006}.\\

High inclination systems have also been found to exhibit X-ray disc winds that are an important aspect in 
understanding the process of accretion/ejection, particularly in LMXBs \citep{Miller_2006,Miller_2015, Ponti2014}. 
Disc-wind outflow signatures have been detected in the form of high ionization Fe absorption lines with large
outflow velocities, in the soft spectral state \citep{Neilsen2009,Ponti2012a}. The fact that these absorption lines are seen only in high inclination sources indicate that these 
absorbers have an equatorial distribution. In both black hole as well as neutron star binaries, equatorial disc-wind outflow 
signatures have been detected only in their soft spectral state \citep{Miller_2006,Neilsen2009,Ponti2012a,Ponti2014}.
 \citet{diaztrigo2006} detected highly ionized Fe absorption lines (Fe XXV and Fe XXVI) in six high inclination dippers. 
Our current study focuses on one such high inclination LMXB, XTE J1710-281, with an aim to bring out important timing and spectral 
features in the source. \\

XTE J1710-281 is a transient eclipsing LMXB that was discovered in 1998 serendipitously, 
with \rxte \citep{Markwardt_1998}. It is a highly inclined binary system viewed edge on with an inclination angle 
i $\sim$75$^\circ$-80$^\circ$ \citep{Younes}. It shows X-ray eclipses every 3.28 hours with an average eclipse duration
of $\sim$420 s \citep{Jain2011}. This LMXB also exhibits X-ray intensity dips.
The discovery of Type 1 thermonuclear bursts indicated that the compact object is a neutron star and the distance to the source was
constrained using the Type-1 bursts to be between 15-20 kpc \citep{Markwardt_2001}.
The spectrum measured with RXTE PCA was consistent with a thermal bremsstrahlung (kT = 14$\pm$3 keV) or a power law (photon index 1.8$\pm$0.1), with an 
absorption column density N$_H$ $<$ 2$\times$10$^{22}$ cm$^{-2}$ \citep{Markwardt_1998}. \citet{Younes} reported spectral changes during dipping intervals
using \xmmn observations in 2004 and compared the partial neutral absorber and ionized absorber models.
Absorption lines due to ionized species like Fe XXV and Fe XXVI observed in other dippers (for example EXO 0748-676: \citealt{Ponti2014},
4U 1323-62 and 4U 1624-490: \citealt{Boirin}) were not detected in their 2004 \xmmn observation. \\

XTE J1710-281 resembles a well studied eclipsing source, EXO 0748-676 in many ways. Both are high inclination LMXBs that show frequent
thermonuclear X-ray bursts and intensity dips in their light curves. EXO 0748-676 has an orbital period of 3.82 hr and
displays pre-eclipse dips that vary significantly within short time-scales \citep{Parmar,RamanPaul17}. \citet{Jain2011} carried out 
extensive eclipse timing studies for XTE J1710-281 and reported two orbital period glitches which are similar to 
the period glitches observed in EXO 0748-676 \citep{Wolff}. \\
 
In this work we have analyzed archival data from \chandra and \suzaku for XTE J1710-281 to study the accretion disc dynamics
in the context of the dip evolution and continuum spectral changes from persistent to dipping intervals. We then compare
it with the eclipsing LMXBs, EXO 0748-676 and AX J1745.6-2901.

\section{Observations and data reduction}
We have analyzed archival X-ray observations of XTE J1710-281 from \chandra and \suzaku observatories, a summary of which, is presented in Table \ref{tab:obs}.
\subsection{\chandra}
XTE J1710-281 was observed with the Chandra X-ray Observatory (CXO, \citealt{Weisskopf}) on the 23rd of 
July and on the 7th of August, 2011. The archival data for each Obs-ID (12468 and 12469) had an exposure of $\sim$75 ks with
the ACIS-S detector using the High Energy Transmission Grating Spectrometer (HETGS; \citealt{Canizares}). The data was processed 
using the \software{ciao} software version 4.7. A level 2 dataset was produced using the \textit{chandra$\_$repro} script. The ACIS 
light curve and spectra were extracted within a source region of 4 arc seconds. The pile-up for \chandra observations
were computed using the web PIMMS version 4.8 \footnote{http://cxc.harvard.edu/toolkit/pimms.jsp}. The ACIS-camera suffered 
from heavy photon pile-up ($\sim$ 65$\%$) for the zeroth order and so the zeroth order data are not used for further analysis for these two Obs-IDs. All the subsequent analysis are done with Chandra first order HETG. The HETG first order grating spectra, arf and rmf were produced during 
the level 2 processing using the \textit{chandra$\_$repro} script. The \textit{combine$\_$grating$\_$spectra} tool was used 
to produce a co-added grating spectrum for the plus and minus first order spectra of the HETG. The corresponding 
arf and rmf files were averaged. The spectra were rebinned to obtain a minimum of 50 counts per bin.

\subsection{\suzaku}
The X-ray Imaging Spectrometer (XIS) aboard \suzaku covers the energy range of 0.2-12 keV \citep{Koyama2007}. The three operational 
XIS detectors: XIS0, XIS1 and XIS2 have been used for this current analysis. The \suzaku Obs-ID 404068010 had a useful exposure
of $\sim$75 kilo-seconds. The XIS detectors were operated in the standard data mode with normal window operation that provided a timing
resolution of 8 s. The unfiltered event files were reprocessed with the CALDB version 20160607 and HEAsoft version
6.16. The XIS event files were checked for pile-up using the ftool `pileest' and were found to be free from photon pile-up 
(less than 0.01\%). The reprocessed event files were used to extract light curves and spectra 
from a circular region of 240 arc seconds around the source centroid. A circular region of the same size far away from the source 
centroid was selected for the background. Corresponding response file was  generated by using the FTOOLS task \textit{xisresp} using the 
CALDB mentioned above. In the case of \suzaku spectra, the spectra from the front illuminated CCDs (XISs 0 and 3), and the back 
illuminated CCD (XIS-1) were fit simultaneously in the energy range of 0.6-9 keV keeping the instrument normalizations free. Due to 
artificial structures in the XIS spectra around the Si and Au edge, the energy range of 1.7-2.3 keV was neglected. The XIS spectra were 
rebinned to contain 100 counts per bin.

\begin{table*}
\centering
\begin{center}
\resizebox{1.1\textwidth}{!}
 {\begin{minipage}{\textwidth}
\begin{tabular}{c @{\extracolsep{\fill}} cccccc}
\hline
Observatory & Obs-ID & Instrument & Date of Observation (MJD) & Exposure (ks) & No. of type-1 X-ray bursts \\
\hline
\hline
Chandra & 12468    & HETG & 23-07-2011 (55765)& 75.1 & 7\\
Chandra & 12469    & HETG & 07-08-2011 (55780) & 75.1 & 5 \\
Suzaku & 404068010 & XIS &  23-03-2010  (55278) & 76.1 & 6\\
\hline
\end{tabular}
 \end{minipage}}
 \caption{Table detailing the different observations used in the current analysis.}
 \label{tab:obs}
 \end{center}
\end{table*}

\section{Timing analysis}
  \subsection{Light curve and average orbital profile}
\begin{figure*}
\centering
 \includegraphics[scale=0.31,angle=-90,trim={1.2cm 1cm 2cm 2cm},clip]{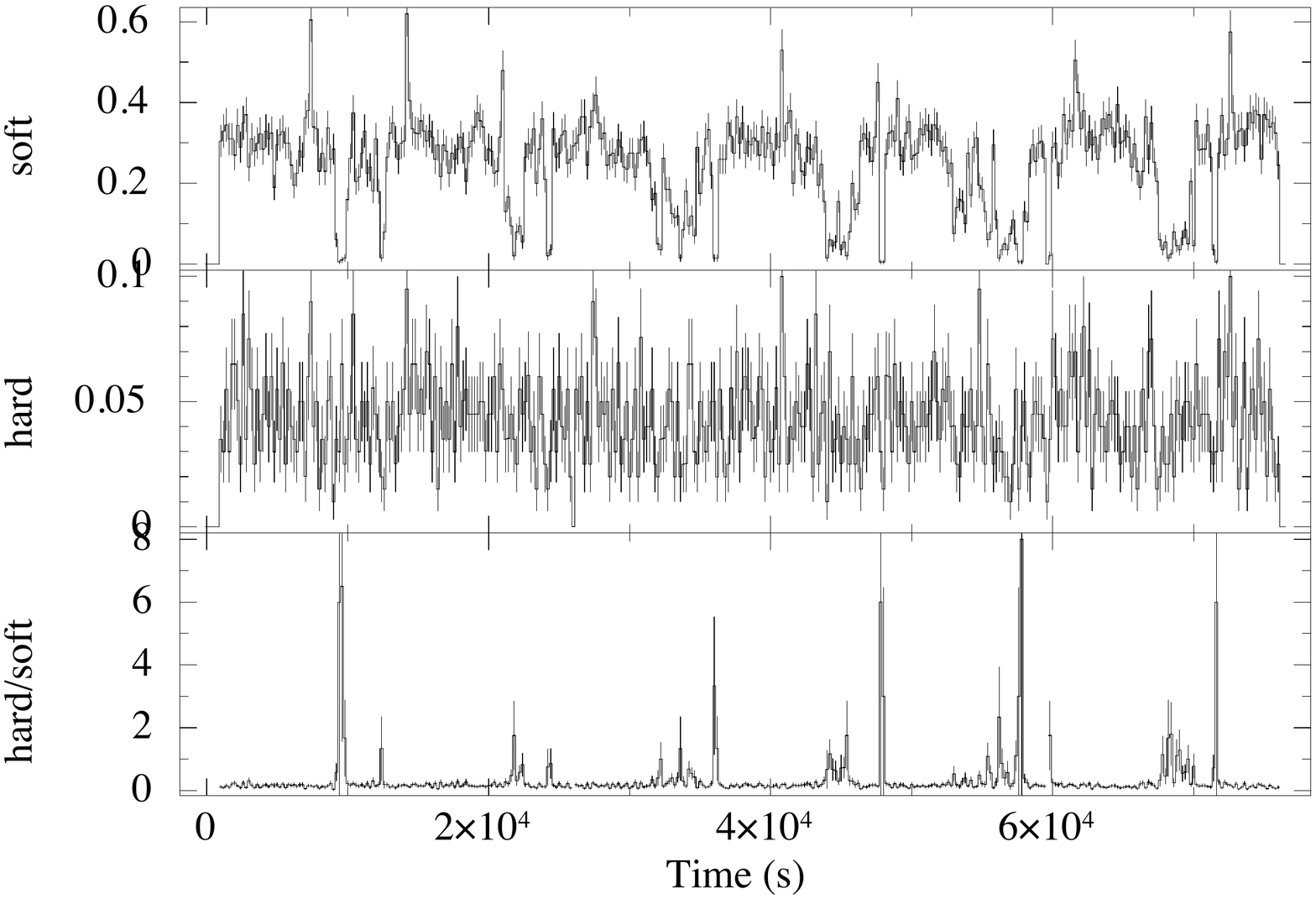} 
 \includegraphics[scale=0.31,angle=-90,trim={1.2cm 1cm 2cm 2cm},clip]{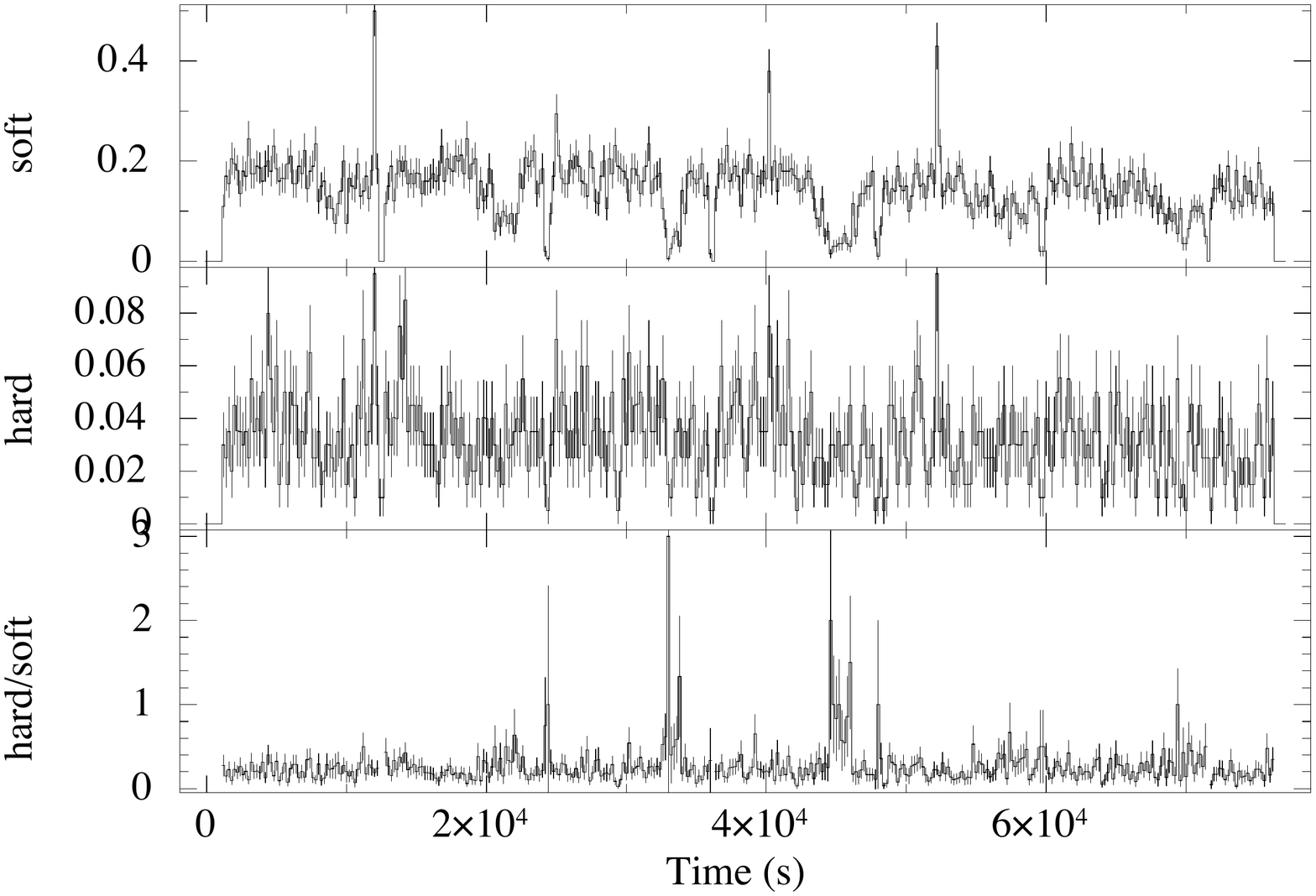} 
 
 \includegraphics[scale=0.31,angle=-90,trim={1.2cm 0 0 0cm},clip]{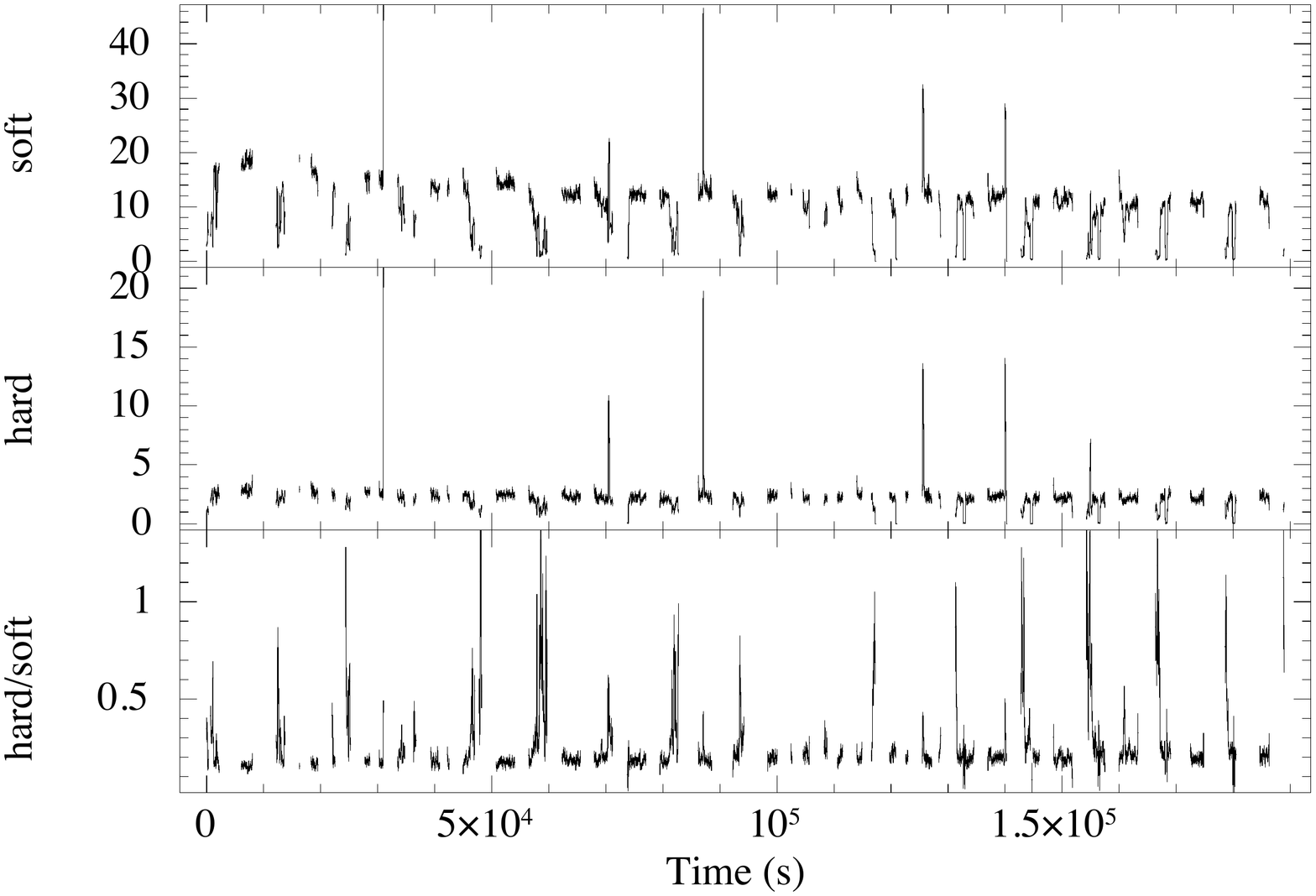}
 \caption{Light curves from the first order HETG \chandra Obs-ID 12468 (top-left), 12469 (top-right) with a bin size of 200 s and 
 \suzaku (bottom) with a bin size of 100 s are shown. The light curves, extracted in two energy bands, soft (0.5-5 keV) and 
 hard (5-10 keV) are shown along with the hardness ratio in the bottom panel for all the three Obs-IDs. An increase in the 
 hardness ratio is observed during dipping.  Type-1 thermonuclear X-ray bursts are also observed in all the light curves.}
 \label{fig:lcs}
\end{figure*}

Since the ACIS zeroth order light curve was heavily piled up, we  extracted HETG first order plus and minus arms 
light curves. Background subtracted and energy dependent (described in detail in section 3.3) first order light curves 
between 0.5 to 10 keV for \chandra Obs-IDs 12468 and 12469 were extracted with a bin size of 1.74 s and are plotted with a 200 s binning
(Figure \ref{fig:lcs} top left and right panels). The light curves display more than 6 orbital cycles with sharp X-ray eclipses. 
X-ray intensity dips are seen to occur at orbital phases prior to the eclipse. The depth and width of the dips are seen to 
vary from one orbital cycle to the next, as described in detail in the next subsection. 
The first order light curves from both the \chandra observations also exhibit 7 and 5 type-1 bursts, respectively, as mentioned in Table \ref{tab:obs}. \\

We also extracted light curve for the \suzaku observation in the 0.5-10 keV energy band with a time bin 
size of 8 s and plotted it with a bin size of 100 seconds (Figure \ref{fig:lcs} bottom panel). The light curve contains large data gaps, with 
incomplete sampling of persistent, dipping and eclipsing intervals. The light curve shows a total of six thermonuclear X-ray bursts 
during this observation. \\

\begin{figure}
\centering
 \includegraphics[scale=0.4,angle=-90,trim={1cm 3cm 0 0cm},clip]{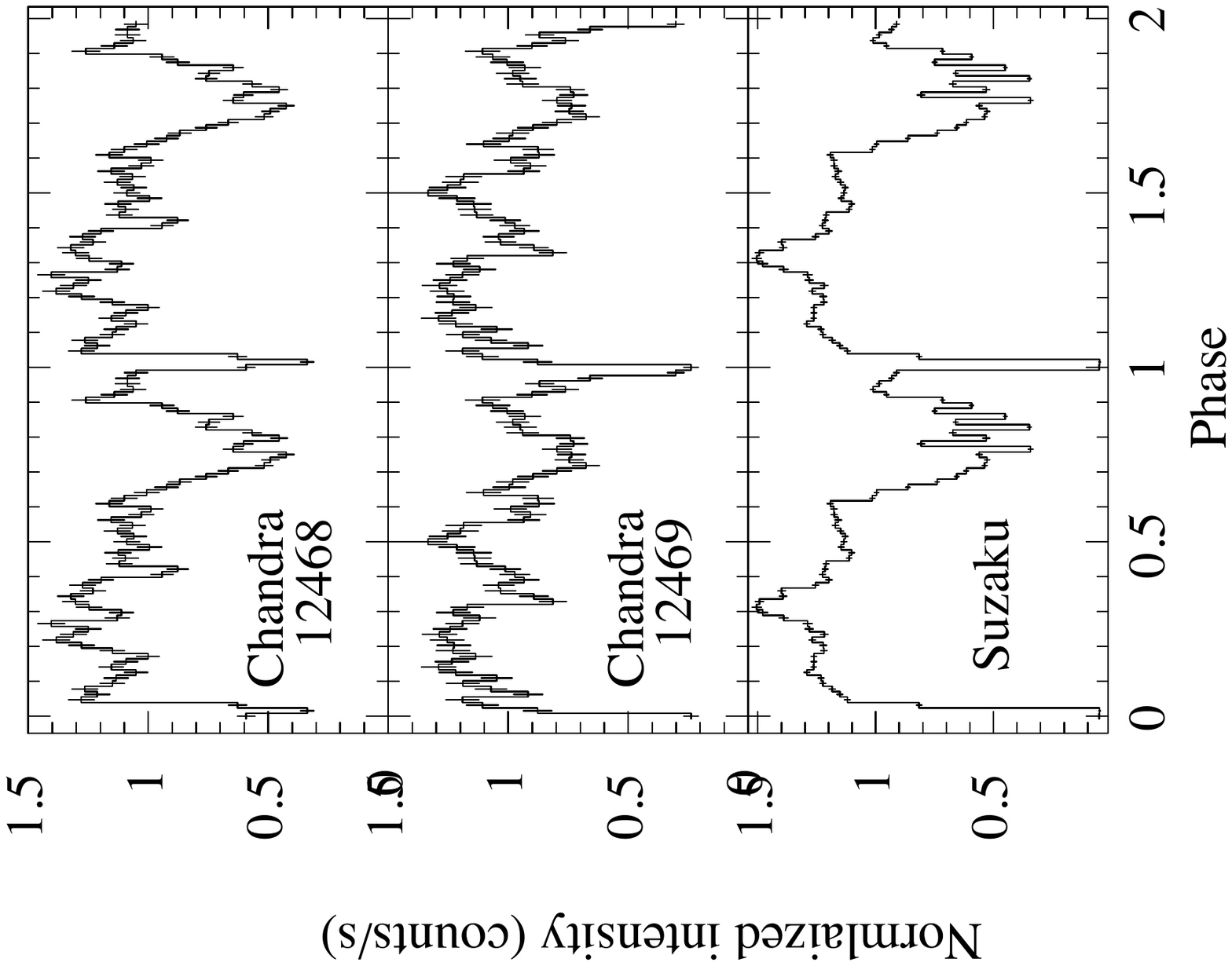}
  \caption{The average orbital profile of the two Chandra observations 12468 (top) and 12469 (middle) folded at 11812.66 s.
 The bottom panel shows the average \suzaku folded pulse profile. All three observations display a clear dip at the orbital phase 
 just before the eclipse.}
 \label{fig:efold}
 \end{figure}

In order to compare the average orbital profile, specially the dip features between the different observations, we folded 
the light curves with the orbital period of 11812.66 s \citep{Jain2011}, at a reference epoch of MJD 55278.29, 55762.79 and 55766.895 
for the \suzaku observation and \chandra Obs-IDs 12468 and 12469, respectively. In order to exclude the thermonuclear bursts, 
time filters were applied to the light curves. These time windows were $\sim$ 100 s in duration. The average orbital profile is shown in 
Figure \ref{fig:efold} with dips covering the phase range 0.6-0.9. The \chandra Obs IDs 12468, 12469 and \suzaku light curves show slightly 
different folded orbital profile features. The duty cycle of a dip is defined as the duration of the dip as a fraction of the total orbital period.
The pre-eclipse dips for the \chandra Obs-ID 12468 and \suzaku have duty cycle of nearly 40\%, while for the \chandra Obs-ID 
12469 it is nearly 30\%. Along with the pre-eclipse dips, the \suzaku profile shows a bump at around phase 0.3, while the 
\chandra Obs-IDs 12468 and 12469 profiles show a less intense secondary dip at phase $\sim$0.4 (duty cycle 10\%).

\subsection{Orbit to orbit evolution of dips}

\begin{figure}
\centering
 
 \includegraphics[scale=0.34,angle=-90, trim={2.2cm 8cm 1cm 10cm},clip]{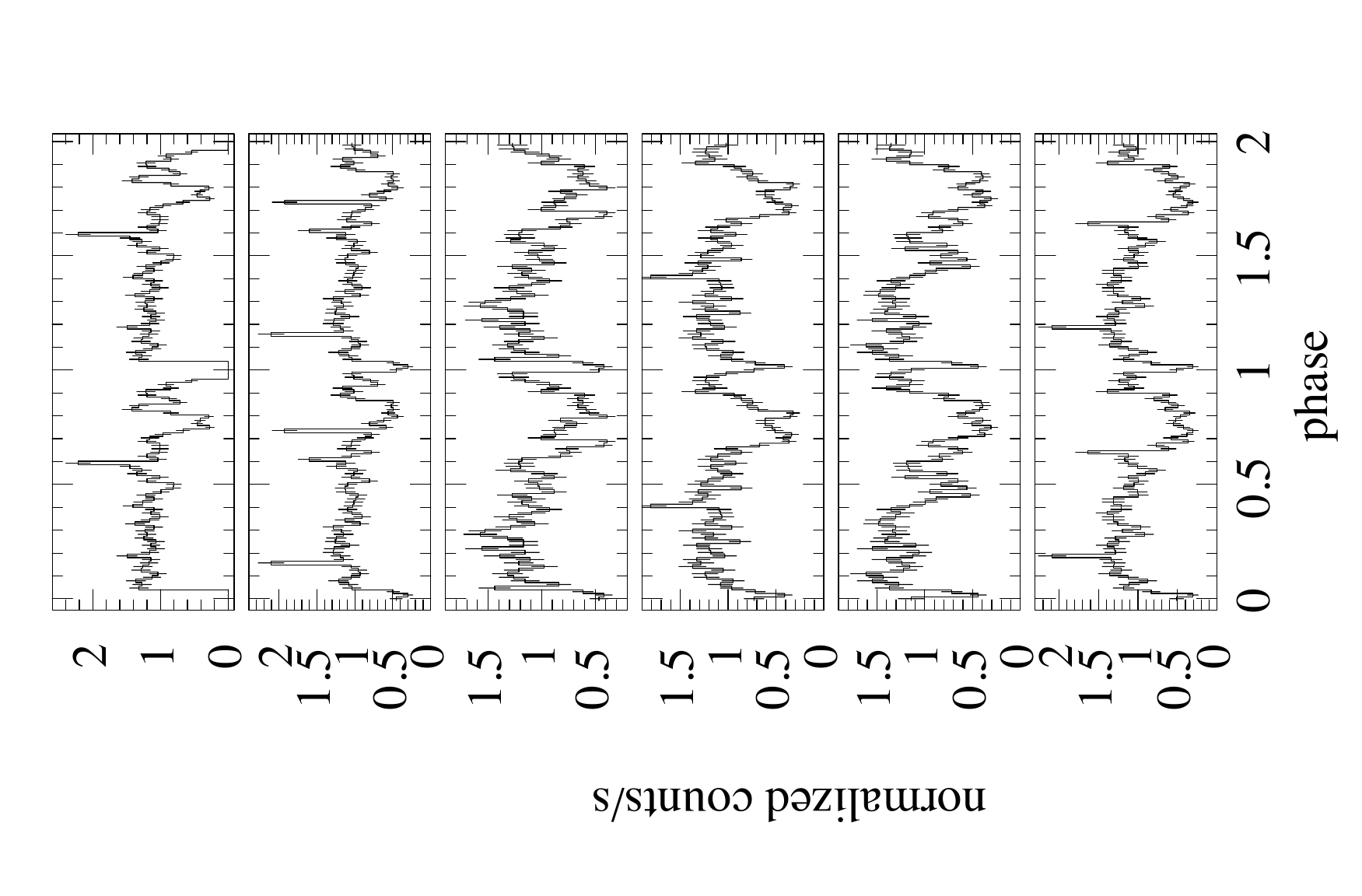}
 \includegraphics[scale=0.34,angle=-90, trim={2.2cm 8cm 1cm 9cm},clip]{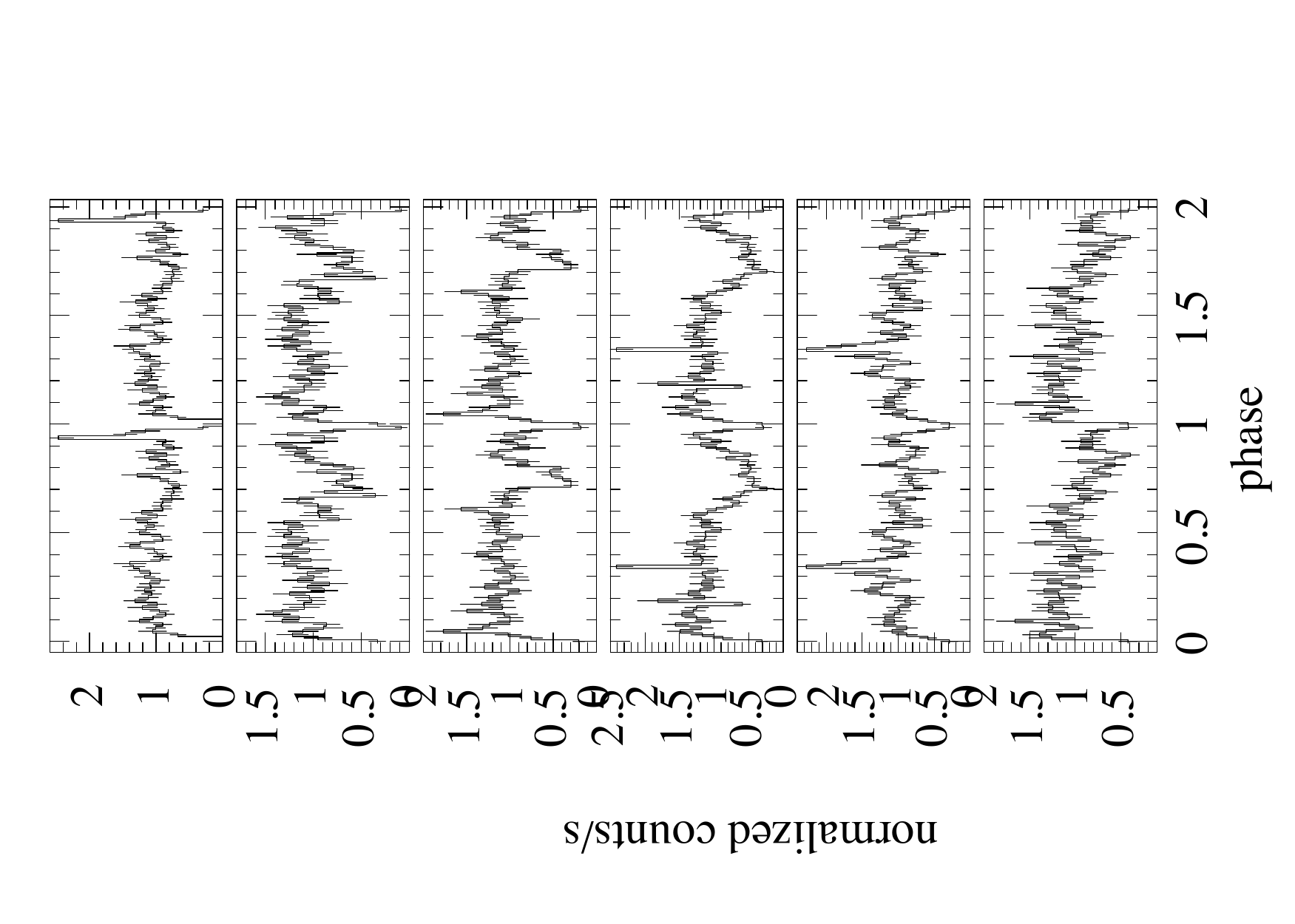}
 
 \caption{The orbit-to-orbit intensity modulations are shown for the Chandra Obs IDs 12468 (left) and 12469 (right) first order
 light curve. A clear evolution of the dip parameters (depth, width and shape) as 
 a function of time, is observed. }
\label{fig:orbtoorb} 
\end{figure}

While examining the \chandra data, an orbit to orbit evolution of dips is seen for the first time in the source (Figure \ref{fig:orbtoorb}).
The evolution of the depth and width in each subsequent orbit is clearly visible for both the observations.
The pre-eclipse dip is observed to be with and without sub-structures in different orbital cycles. An earlier study of the dips made by
\citet{Younes} in 2004 using \xmmn also reported presence of narrow absorption dips 
in one single orbital cycle alone.
For both the observations, the duty cycle of the dips range from 
about 10\% to nearly 50\%, possibly indicating variable absorber azimuthal distribution above the disc plane. The depth of 
the dips vary between 20\%-90\%, signifying variable absorber column density along our line-of-sight between successive 
orbital cycles. Unlike the light curves from the \chandra observations, the \suzaku light curve has data gaps. We cannot 
therefore investigate the dips in the light curve in successive orbits.

\subsection{Energy dependence of dips}
To investigate the energy dependence of the dips, we extracted background subtracted light curves in two energy bands: 0.5-5 keV and 
5-10 keV for all the observations. Figure \ref{fig:lcs} shows the energy dependent intensity variations in the light curves for
the three observations. The soft X-ray light curve shows more prominent dips as 
compared to the hard-X-ray band. As a result, the hardness ratio increases during the X-ray 
dips, with the deeper troughs showing an increased hardness ratio compared to the shallower one. 
Although the \suzaku observation has periodic large data gaps, a periodic increase in the hardness ratio with interval same as 
the orbital period indicate it to be due to the dips in the same orbital phase range as the \chandra light curves. 
The increase in hardness ratio during dipping intervals has been observed before by \citet{Younes} using \xmmn observations 
for this source, and for other dipping sources like 4U 1254-690, 4U 1323-62, 4U 2127+11 etc.
\citep{Ioannou_2002,Boirin,diaz_2009}.

 \section{Spectral Analysis}
 \subsection{Time averaged spectroscopy}
  
 We performed spectroscopy of XTE 1710-281 averaged over the entire stretch of the observations for the \chandra and \suzaku data.
 Spectral fitting was performed using \textit{XSPEC} \citep{arnaud_1996} version 12.8.1. The photo-electric cross-sections of \citet{Wilms_2000} 
have been used to model the line-of-sight absorption by neutral gas with solar abundances (the `\textit{TBabs}'
\textit{XSPEC} model). The cross section for X-ray absorption by the ISM is computed as the sum of the cross sections for X-ray 
absorption due to the gas-phase ISM, the grain-phase ISM, and the molecules in the ISM. \\

\begin{figure*}
 \centering
 \includegraphics[scale=0.33,angle=-90,trim={1.8cm 0cm 1cm 3cm},clip]{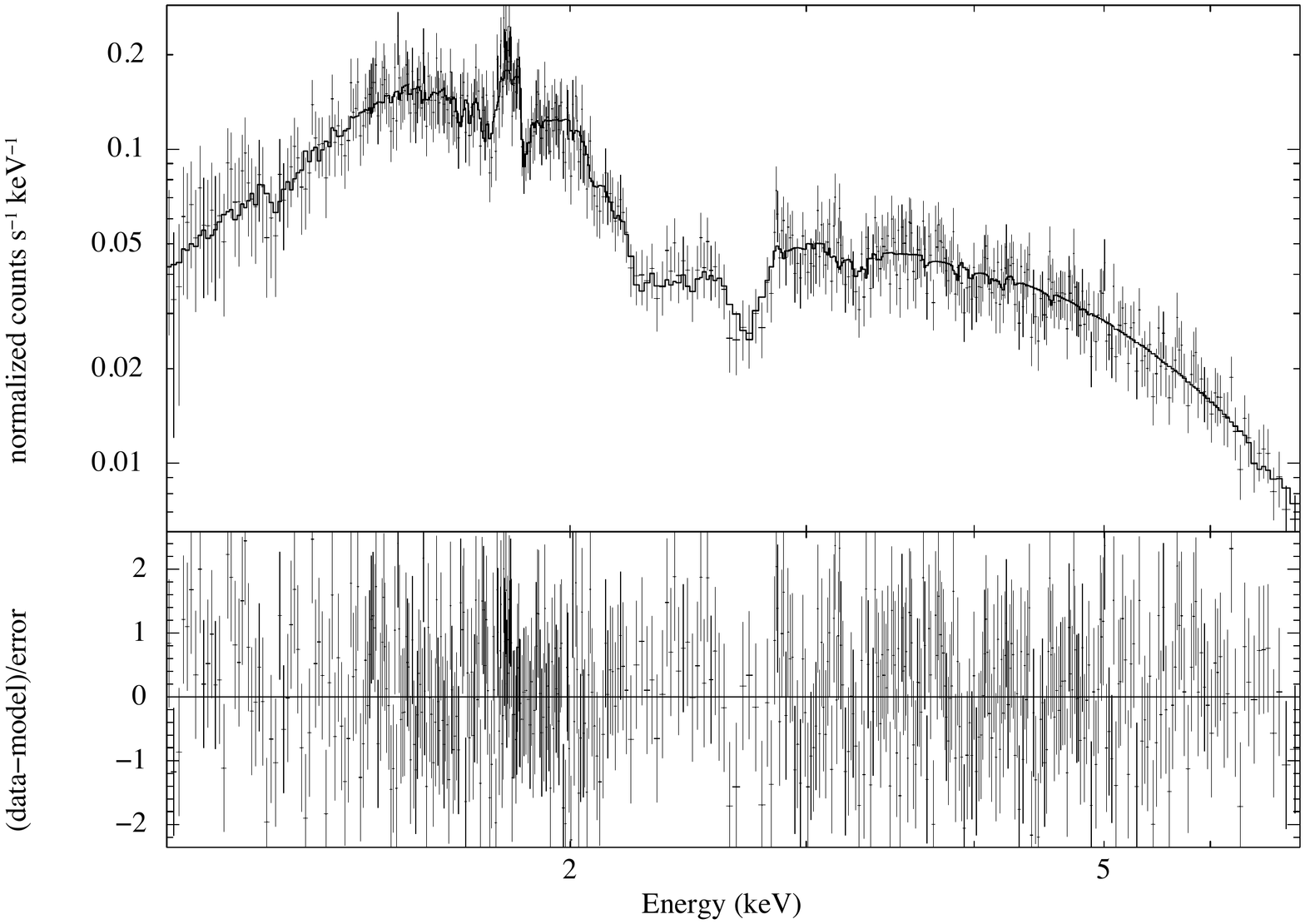}
   \includegraphics[scale=0.33,angle=-90,trim={1.8cm 1.2cm 1.3cm 3cm},clip]{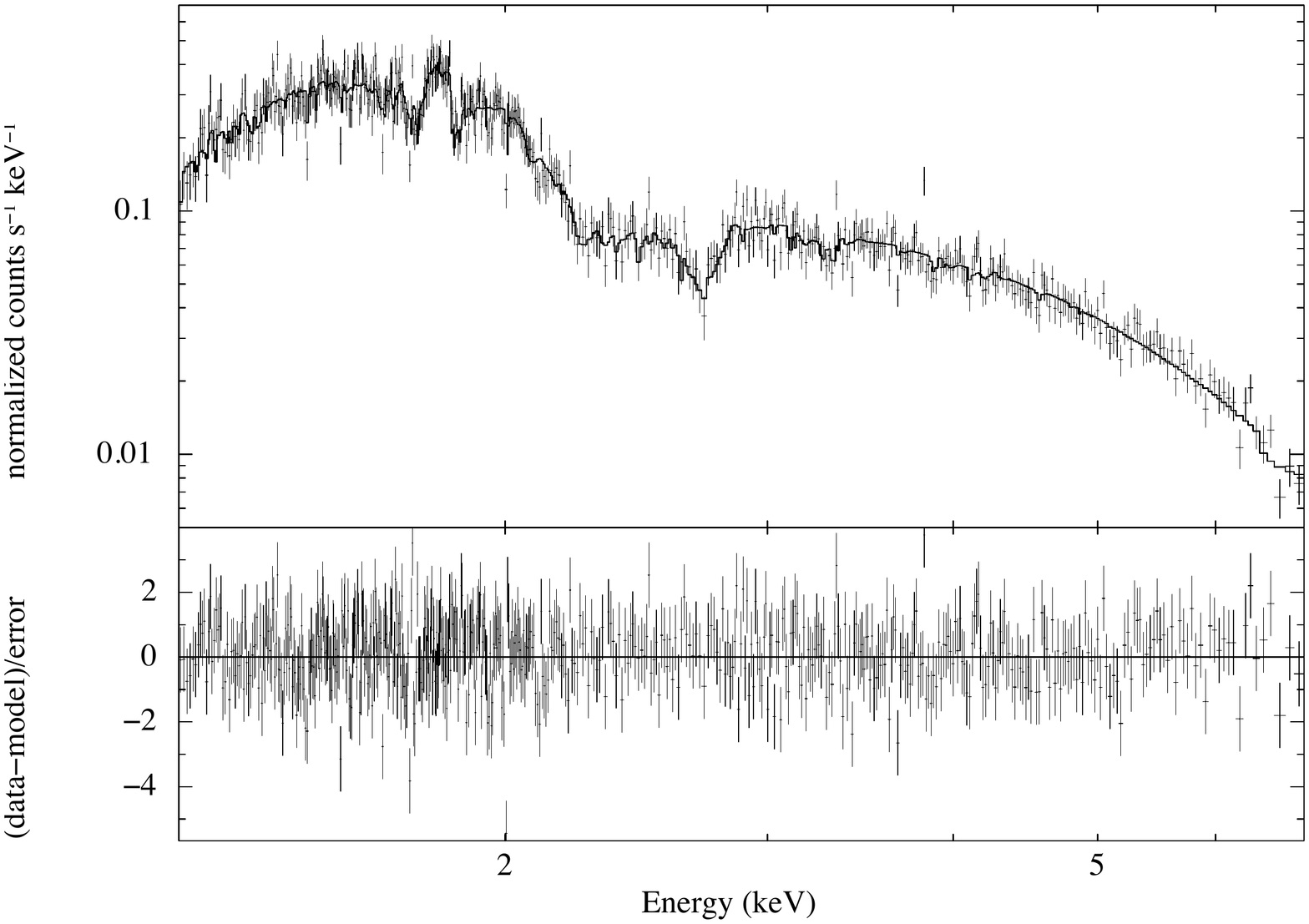} 
\includegraphics[scale=0.3,angle=-90,trim={0cm 2cm 1.3cm 4cm},clip]{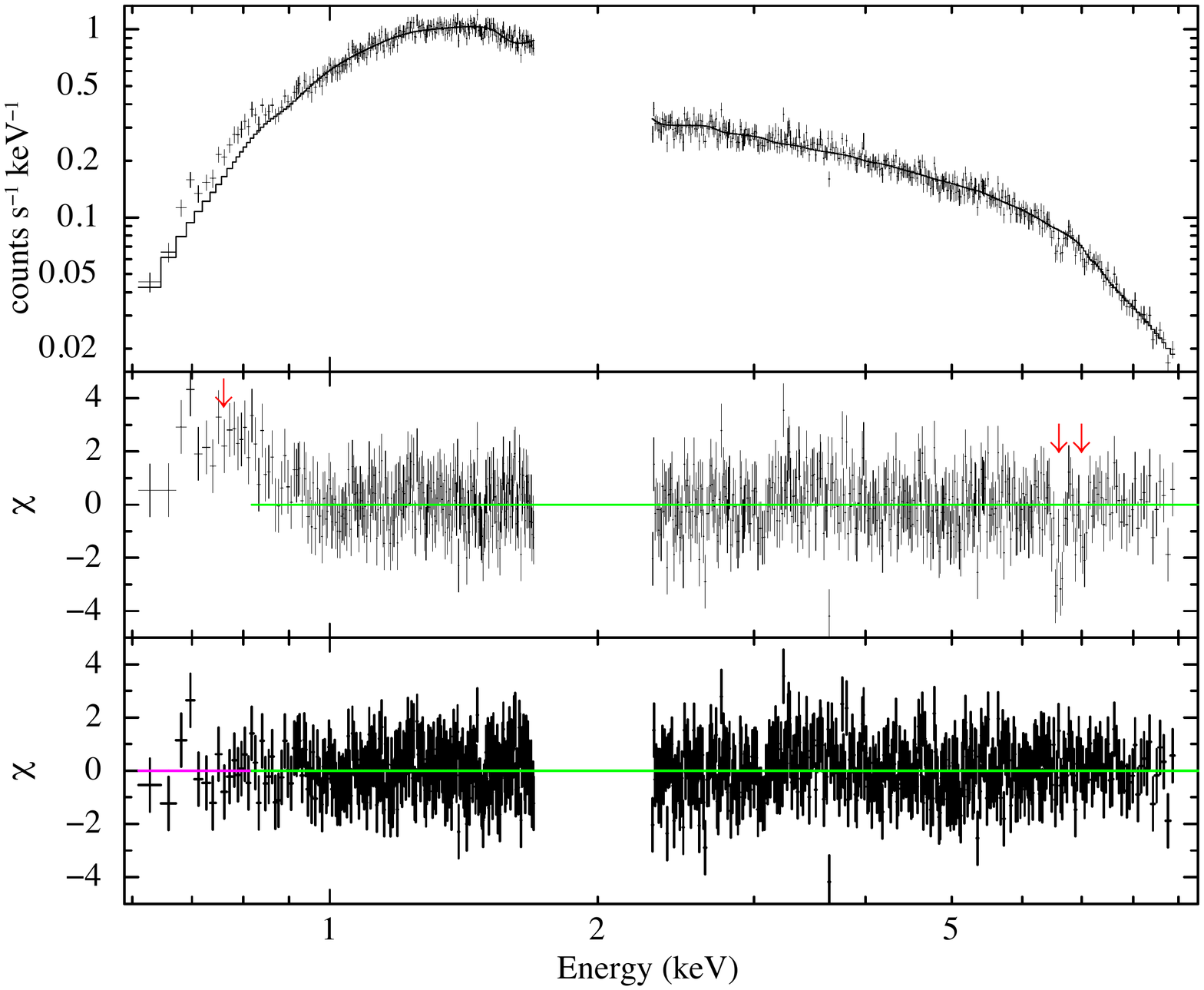} 
    \caption{The time averaged best fit spectra for \chandra Obs-ID 12468 (top-left), Obs-ID 12469 (top-right) and \suzaku (bottom) are shown.  For the \chandra Obs-ID 12468 spectrum, the plot has been re-binned for visual clarity. 
    In the \suzaku spectrum, the residuals are indicated without (top residual panel) and with (bottom residual panel) 
    the 0.72 keV, 6.59 keV and 7.01 keV lines modelled. }
    \label{fig:avgspec}
 \end{figure*}

  At first all the spectra were fitted with a powerlaw with a single absorption component accounting 
 for the foreground absorption which includes the line of sight interstellar absorption as well as absorption local to the source. 
Fitting a simple power-law model to the \chandra observation 12468 resulted in a reduced chi-squared value of $\sim$0.65. 
We therefore re-binned the spectrum to at least 1 count per bin and applied C-statistic which is appropriate for the quality of the data. 
The spectral fit with an absorbed power-law provided an acceptable fit with a C-stat/dof of 2305/3841 (Figure \ref{fig:avgspec} top left panel).\\ 

 In the case of the other \chandra (Obs-ID 12469)  and \suzaku observation, the fit with a simple absorbed power-law 
  proved to be unsatisfactory with large systematic residuals particularly in the energy  range 1.2 - 2.0 keV.
   To account for this we added another absorption component in the form of a partial covering absorber. This can be explained due to  absorption by local structures in the accretion disc (causing the pre-eclipse dips). We tried to fit the spectra with both a partial covering 
 neutral absorber (\textit{XSPEC} model `\textit{pcfabs}') and partial covering ionized absorber (\textit{XSPEC} model 
 `\textit{zxipcf}', \citealt{Reeves_2008}). 
  For the \chandra Obs-ID 12469, the addition of partial covering neutral absorber improved the $\chi^2$ by 12.1 ($\Delta$d.o.f. of 2) and the addition of partial covering partially ionized absorber improved the $\chi^2$ by 79.5 ($\Delta$d.o.f. of 3).
  In the case of the \suzaku observation, the reduction in $\chi^2$ obtained by adding the partial covering neutral and ionized absorber models, was 37 ($\Delta$d.o.f. of 2) and 50 ($\Delta$d.o.f. of 3), respectively.
 The fit with the ionized absorber model proved to be better both in terms of absence of systematic 
 structures in the residuals and a lower value of reduced $\chi^{2}$. A power-law model with a partially ionized partial covering 
 absorber model was therefore chosen as the best fit model and all subsequent analysis were performed with this. \\
 
 
 In addition to the modelling of the continuum, we detected several absorption lines and edge features in the \chandra and 
 \suzaku spectra. 
 In the time-averaged \suzaku spectrum, a broad emission feature is observed at $\sim$0.72 keV with an
 equivalent width  93.03$\pm$66.98 eV. 
 This is probably a complex due to unresolved Fe L 
 shell lines. Two additional weak absorption lines are observed at 6.60$\pm$0.03 keV and 7.01$\pm$0.03 keV. These two absorption 
 lines were modeled using gaussian absorption model. 
The addition of the 0.72 keV feature improved the $\chi^2$ by 31, with the addition of 3 new parameters for a $\Delta$d.o.f of 3. 
Further addition of the two absorption features at 6.6 keV and 7.01 keV progressively improved the $\chi^2$ by 23 ($\Delta$d.o.f of 3), and 7 ($\Delta$d.o.f of 3), respectively.  The 6.6 keV line is detected with an equivalent width of 4.28$\pm$0.47 eV. The 7.01 keV absorption line, being not statistically significant is detected with an equivalent width of only $\sim$0.58$\pm$0.49 eV.
 The best fits for the time-averaged spectra after adding all the components gives a reduced $\chi^2$ of 1.10 for 566 d.o.f. for 
 the \chandra Obs-ID 12469 spectrum and 0.95 for 599 d.o.f for the \suzaku spectrum. \\

 Although we do not detect the $\sim$6.6 keV feature 
in the \chandra HETG spectrum, we cannot rule out their presence in the corresponding spectra. This is because, for the count rate of XTE J1710-281, these lines would be below the detection threshold of HETG considering its lower effective area w.r.t the XIS. In order to examine if such a feature would be detectable in the \chandra spectrum, a gaussian absorption feature (the `gabs' model in XSPEC) was introduced at 6.60 keV, with the line energy and width fixed to the parameter values obtained in the Suzaku spectra. This model is, however, not statistically preferred ($\Delta \chi^{2}$ of 0.11 for $\Delta$d.o.f of 3) as an additional component over the continuum. We therefore do not include it in the spectral fit. \\


 Table \ref{tab:avgspec} shows the best fit parameters along with the fluxes measured for the \chandra and \suzaku observations.
 Figure \ref{fig:avgspec} shows the best-fit average HETG first order spectra for Chandra Obs ID 12469, and the XIS0 \suzaku spectra.
 The derived absorbed fluxes (0.5-8 keV) for the \chandra (Obs-ID 12468 and 12469) and the \suzaku observations in order are 5.14$\pm$0.20$\times$10$^{-11}$ ergs s$^{-1}$ cm$^{2}$, 7.94$\pm$0.10$\times$10$^{-11}$ ergs s$^{-1}$ cm$^{2}$ and 4.10$\pm$0.07$\times$10$^{-11}$ ergs s$^{-1}$ cm$^{2}$, respectively. 
 The \chandra Obs-ID 12469 shows a steeper power law index and larger flux compared to the \chandra Obs-ID 12468 and \suzaku observation.\\

\begin{table*}
\centering
\begin{center}
\begin{tabular}{ @{\extracolsep{\fill}} ccccc}
\hline
Model Parameter & Model & \chandra Obs-ID 12468 &\chandra Obs-ID 12469 & \suzaku \\
\hline
&&&&\\
nH$^{fore}$ (10$^{22}$cm$^{-2}$) & \textbf{TBabs}  & 0.21$\pm$0.1 & 0.72$\pm$0.1 & 0.51$\pm$0.03 \\
nH (10$^{22}$cm$^{-2}$) & \textbf{zxipcf}   &   &45.3$^{-10.4}_{+9.2}$ & 45.96$\pm10.1$ \\
log($\xi$) & \textbf{zxipcf}  &  & 3.03$\pm$0.1  & 1.91$\pm$0.34 \\
Cov. f & \textbf{zxipcf}    & & 0.56$\pm$0.1 & 0.45$\pm$0.05 \\
Power law index & \textbf{po}  &  1.54$\pm$0.2 &2.45$\pm$0.1  &1.98$\pm$0.06 \\
E$_{gabs}$ (keV) & 	\textbf{gabs}   & &    &6.60 $\pm$0.03\\
EW (eV) & \textbf{gabs} & & & 4.28$\pm$0.47 \\
E$_{gabs}$ (keV) & 	\textbf{gabs}    &  &   &7.01$\pm$0.03\\
EW (eV) & \textbf{gabs} & & & 0.58$\pm$0.49 \\
E$_{gaus}$ (keV) & 	\textbf{gaus}    &   &   & 0.72 $\pm$ 0.01\\
EW (eV)  & \textbf{gaus} & &   & 93.03$\pm$66.98\\
 \hline
 red. $\chi^{2}$& & C-stat/dof: 2305/3841   & 1.10 for 566 d.o.f. &0.95 for 599 d.o.f\\
 &&& &\\
 Flux (ergs/cm$^{2}$/s)$^{a}$&  & 5.14$\pm$0.20 $\times10^{-11}$ & 7.94$\pm$0.10 $\times10^{-11}$ & 4.10$\pm$0.07 $\times10^{-11}$ \\
 \hline
\end{tabular}
\caption{Best fit spectral parameters for time-averaged \chandra (Obs-IDs 12468 and 12469) HETG spectra and
 \suzaku XIS0 spectra. The errors are quoted at 90$\%$ confidence interval.$^{a}$ Flux is measured from 0.5-8 keV}
\label{tab:avgspec}
 \end{center}
\end{table*}

\subsection{Intensity resolved spectroscopy}
In order to probe the changes in the spectrum from the persistent and dipping intervals, we carried out an intensity 
resolved spectral analysis of the \chandra and \suzaku observations. In order to perform this exercise, we first excluded the 
intervals corresponding to thermonuclear bursts and eclipses. We then defined persistent and dipping intervals based on the 
count rate from the light curves of the corresponding observations.
For the \chandra observations the persistent emission is seen to have an average count rate of 0.25 counts/s (Figure \ref{fig:lcs}).
Time intervals that had intensities below 75\% of the average persistent count rate were considered as dipping intervals.
 We then extracted spectra corresponding to those intervals. The \chandra intensity resolved spectra were grouped to contain 50 counts per energy bin. We did not carry out intensity resolved spectral analysis for the \chandra Obs-ID 12468 due to statistical limitations.\\

Since the \suzaku light curve contained incompletely sampled dips, it was not possible to identify dipping time intervals from the 
light curve. Therefore a count rate cut of 75\% of the average persistent count rate (about 2 counts/s), similar to what was used for 
the \chandra analysis, was adopted to identify the dipping intervals. The corresponding spectra for the persistent and dipping 
intervals were extracted. A different grouping scheme was adopted for the persistent (minimum of 120 counts per bin) and dipping 
(minimum of 100 counts per bin) intervals, as compared to the time-averaged \suzaku spectrum to have comparable statistics. 
The same background and response files that were used to fit the persistent spectra were taken for the fits.\\ 
  
  
\begin{figure}
 \centering
\includegraphics[scale=0.3,angle=0,trim={0.5cm 1.5cm 1.0cm 2.5cm},clip]{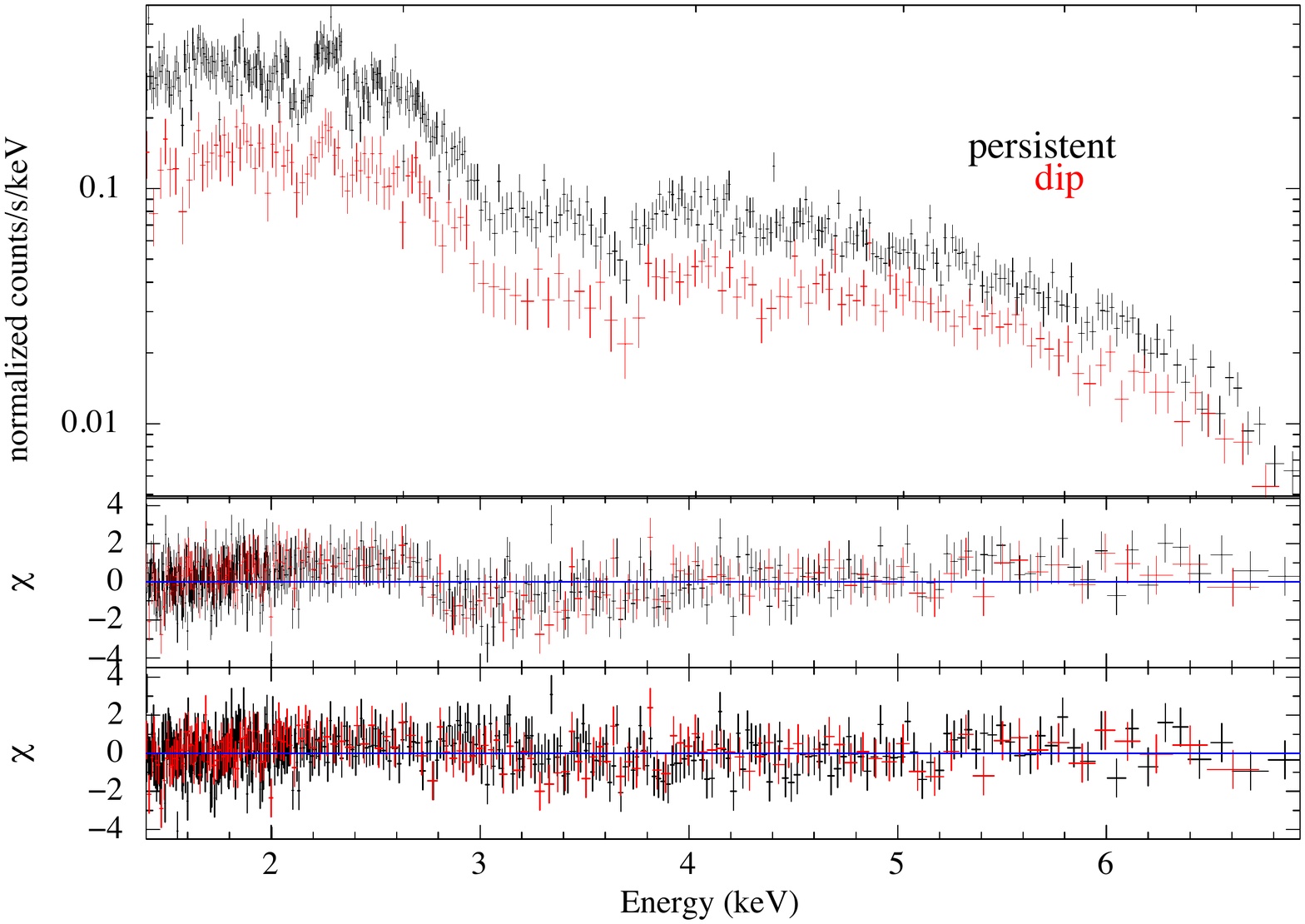} 
  \includegraphics[scale=0.31,angle=-90,trim={2.5cm 0.6cm 1cm 0cm},clip]{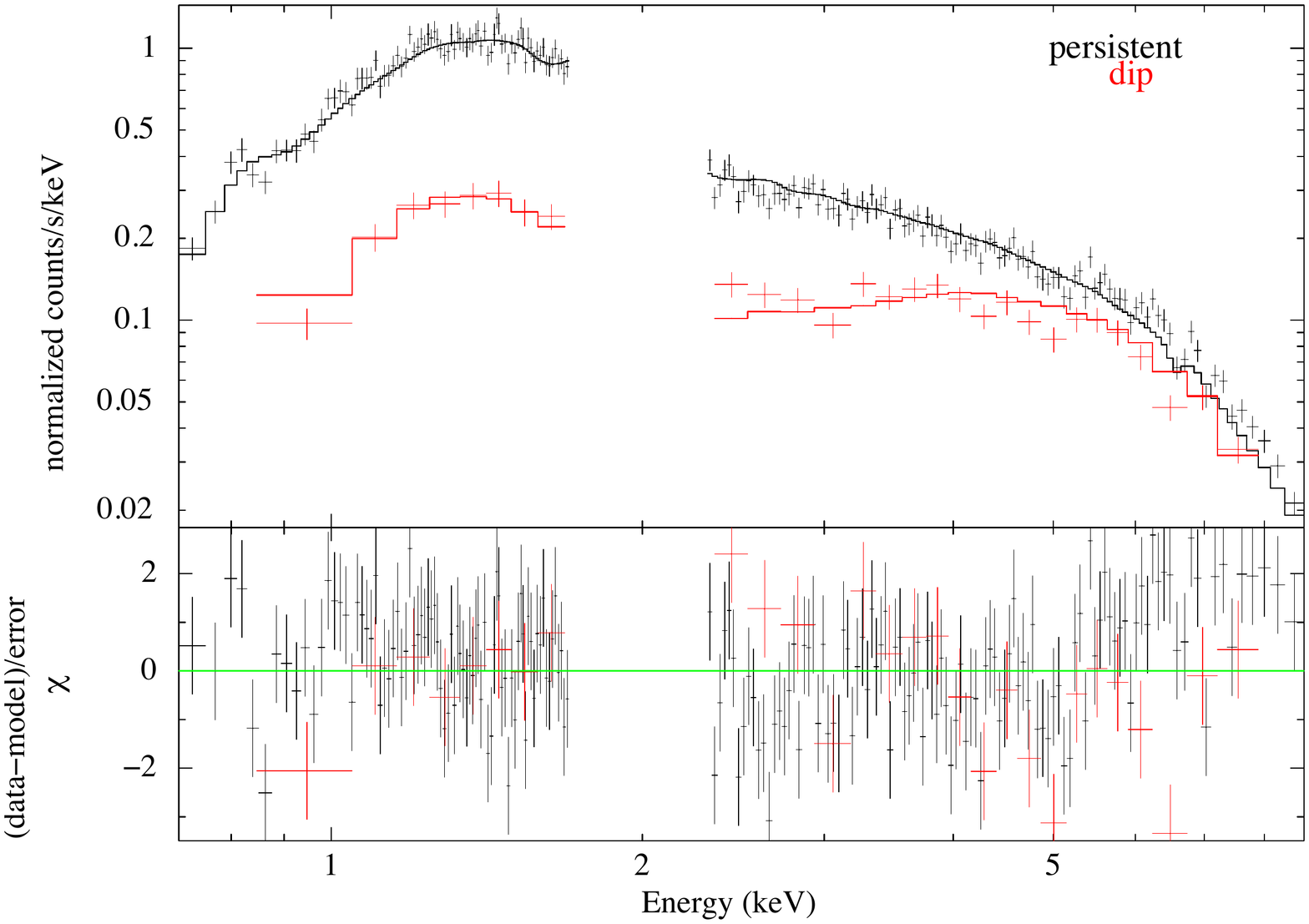} 
  \caption{Intensity resolved spectra for the persistent and dip epochs are fit simultaneously for \chandra Obs-ID 12469 (top) 
  and \suzaku (bottom) observations. The two panels showing the residuals in the \chandra spectra, correspond to the edge at 2.8 keV 
  left unmodelled (top residual plot) and modelled (bottom residual plot).  The best fit spectrum for only the XIS0 is shown for 
  the \suzaku observation.}
  \label{fig:intres}
 \end{figure}


\begin{table*}
\centering
\begin{tabular}{ @{\extracolsep{\fill}} p{2cm} p{1cm}  p{1.4cm} p{1.2cm} p{1.5cm} p{1.4cm} p{1.4cm} p{1cm} }
\hline
 Parameter & Model  & \multicolumn{3}{c}{\chandra Obs-ID 12469} & \multicolumn{3}{c}{\suzaku} \\        
 & & Persistent &  & Dip & Persistent& & Dip\\
 \hline
 &&&&&&\\
   nH$^{fore}$ (10$^{22}$cm$^{-2}$) & \textbf{TBabs} &   &0.72$^a$ (fixed)  & &0.23$\pm$0.01  &  & 0.17$\pm$0.06\\
   nH  (10$^{22}$cm$^{-2}$) & \textbf{zxipcf} &  12.9$^{-9.1}_{+17.2}$ & &65.6$\pm$25.2  & 0.36$\pm$0.15 & & 14.9$\pm$1.8\\
   log($\xi$) & \textbf{zxipcf} &  3.43$^{-0.1}_{+0.57}$ & & 1.85$^{+0.2}_{-1.0}$   & 2.6$\pm$0.15 & & 1.45$\pm$0.1\\
   Cov. f & \textbf{zxipcf} &   0.65$\pm$0.4 &  & 0.5$\pm$0.02 &   0.67$\pm$0.3 &  &0.89$\pm$0.01\\
   Power law index & \textbf{po} &  & 2.57$\pm$0.04 & & &1.79$\pm$0.02 & \\
 
  
  
%
%
%
%
     E(keV) & \textbf{edge} &   & 2.8$\pm{0.2}$ & & & &\\ 
     Max Tau & \textbf{edge} &  & 0.45$\pm$0.10   &   &  & &\\
   \hline
     red. $\chi^{2}$&  & &0.92 for 581 d.o.f. &  & &1.13 for 375 d.o.f &\\
 \hline
\end{tabular}
\caption{Best fit spectral parameters for intensity resolved \chandra (Obs-ID 12469) HETG spectra and \suzaku 
XIS0 spectra. The errors are quoted at 90$\%$ confidence interval. $^{a}$Frozen to the time-averaged spectral parameter value.}
 \label{tab:intres}
\end{table*}

We consider the cause of the dipping activity to be absorption due to the passage of the vertical 
structures obscuring our view to the source \citep{White_swank, diaztrigo2006}. In this case we may assume that the 
underlying continuum spectrum does not change during the duration of the observation, and the spectral changes during the dipping 
intervals are solely due to the passage of this absorbing material along our line of sight. In order to demonstrate this we fit 
the persistent and the dipping spectra simultaneously linking the parameters of the continuum emission, \textit{i.e.} the 
powerlaw component. The absorption components are left free. The persistent emission is not used for constraining the ionized 
absorber properties.
We fix the energy and width of the emission and absorption lines and edges in the  \chandra and \suzaku spectra to 
the corresponding values obtained in the average spectra.\\
 
In the case of the intensity resolved spectral fit for the \chandra Obs-ID 12469, the line-of-sight neutral hydrogen 
column density was fixed to the time-average value. This was because the errors obtained on this parameter, if left free for each 
component, was very large. The \chandra intensity resolved spectrum showed a residual, resembling an edge around 2.8$\pm$0.2 keV, 
 which we have modeled using an edge model. This edge, possibly indicating the presence of S XVI, removes the residual and improves the fit ($\Delta\chi^{2}$ 356.49 and $\Delta$dof 3).
  The addition of this component in the time average spectrum only shows a small improvement in $\chi^2$ ($\Delta\chi^2$ of 5), making it a weak detection.
We observe an increase in the column density of the ionized absorber and a decrease in the ionization state from 
the persistent to dipping spectra. A similar change in the ionized absorber parameters is observed for the intensity 
resolved \suzaku spectra as well. The column density of the ionized absorber increases from 0.36$\pm$0.15$\times$10$^{22}$cm$^{-2}$
to 14.9$\pm$1.8$\times$10$^{22}$cm$^{-2}$ and the ionization parameter shows a decrease from 2.6$\pm$0.2 to 1.45$\pm$0.1 as we go from
the persistent to the dipping spectra. We fixed the 6.60 keV and 7.01 keV absorption line parameters, 
and the 0.72 keV broad emission line parameters to the time-averaged spectral parameter values. The best fit spectra for the
intensity resolved \chandra and \suzaku observations are shown in Figure \ref{fig:intres}, and Table \ref{tab:intres} 
shows the best fit spectral parameters.\\

\section{Discussion}
We have carried out spectro-temporal analysis of the dipping LMXB XTE J1710-281 using archival \chandra and \suzaku observations.
We have observed evolution in the dip strength, duration and shape between successive orbital cycles, indicating 
evolution of the accretion disc structures in the timescale of hours. We have also detected highly ionized Fe species in 
the \suzaku spectra, which indicate signature of disc winds as observed in other dippers. 
The dip spectrum in \chandra and \suzaku observations is found to be best described by a power law component with a spectral index of 
$\Gamma$$\sim$2 with partial covering by a partially ionized absorber. 

\subsection{Evolution of the dips}

High inclination LMXB sources exhibit intensity dips, which are usually characterized by increased spectral hardness \citep{diaz_2009}.
These dips have been observed in some total eclipsing sources like EXO 0748-676, XTE J1710-281 and also in 
some partial eclipsing sources like 4U 2129+12 \citep{Chou}. Recently \citet{Galloway_2016} discovered, in the LMXB source Aql X-1, 
intermittent episodes of dipping, similar to those observed in high inclination binaries. Detailed characterization of dip parameters 
like the dip orbital phase, width and strength have been performed for many dippers, to extract information of orbital period and 
also dynamics of the accretion disc structures \citep{Chou,Chou_2001,Hu_2008}. Since dips are caused by absorption in the 
accretion disc structure, they are not exactly phase-locked, but they show phase jitters (eg. $\pm$0.05 cycle for 4U 1915-05,
\citealt{Chou_2001}). In the case of XTE J1710-281 (\citealt{Younes} and the current work), 
on average, the dips are observed pre-dominantly at pre-eclipse orbital phases between the orbital phase range of 0.6-0.9. 
In the system EXO 0748-676, dips have also been seen at other orbital phases, even just after the eclipse \citep{RamanPaul17}.
 A very interesting outcome of this current study of dips using Chandra observations is the evolving morphology from one orbital 
cycle to the next. The dip depth, width as well as shape (in terms of the number of narrow features present) are observed to vary. 
Irradiation-induced vertical structures \citep{deJong,Pringle}, present usually at the outer accretion disc are 
azimuthally distributed in an asymmetric manner \citep{Hakala}. They are responsible for preferentially absorbing the primary X-ray 
photons at certain orbital epochs as was also observed in EXO 0748-676 \citep{Parmar}.
The orbit-to-orbit dip profiles as seen in the \chandra light curves have different features, some are broad, some deep and some 
even have narrow features, similar to what was observed by \citet{Younes} in the \xmmn light curve. 
Nevertheless, the presence of varying dip intensities in successive orbital cycles in \chandra indicates not just any structure, but
an \textit{evolving} one. Warped discs also present a very physically compelling solution to explain different X-ray modulations in LMXBs
\citep{Ogilvie}. These warps can develop at various disc-radii and can modulate the X-ray emission at certain orbital epochs.
Another possible explanation for a hour-time-scale-varying dip morphology could be the variable mass accretion splashes at the 
stream-impact point on the outer accretion disc \citep{White_swank}. At these locations, the disc-extent increases vertically leading 
to temporary structures that block the primary X-ray photons from reaching the observer. 

\subsection{Detection of ionized species of Fe }

The broad 0.72 keV emission line detected in the \suzaku spectrum, can be identified with an Fe L blend.   
It has an equivalent width of 93.03$\pm$66.98 eV. Owing to the fact that XTE J1710-281 is a relatively faint source,
the \chandra spectrum does not have enough photons below 1 keV and was therefore not able to detect a line of 
similar strength. Line emission in the spectral energy range 0.7-2.0 keV, associated with the
Fe L-shell ions, is understood to be a dominant spectral component arising in coronal plasmas \citep{liedahl}. Only 
ionized Fe ions (Fe XII and beyond) produce Fe L transitions that fall in the energy range 0.7-2 keV. 
In addition, two narrow, weak  absorption lines are observed at 6.60 keV and 7.01 keV in the \suzaku spectrum (see Table \ref{tab:avgspec}).
The absorption line at $\sim$6.6$\pm$0.03 keV is most likely a blend of Fe XIX to Fe XXV lines \citep{diaztrigo2006, Ponti2014}.
 Associating it to the nearest Fe XX-XXII transition at $\sim$6.58 keV \citep{Bailon}, results in a blue-shifted velocity of 789 km/s (see for example as computed in \citealt{King2014}). Such blue-shifted velocities correspond to outflows such as disk wind as observed in several other disk accreting high inclination sources. The 7.01+/-0.03 keV is, however, only marginally detected and we do not compute the velocities for this line. Such features at $\sim$6.6 keV have been observed in other NS LMXBs, for example XB 1323-619, XB 1916-053
\citep{diaztrigo2006, Boirin}. 
In their 2004 \xmmn observation of XTE J1710-281, \citet{Younes} did not detect absorption lines due to ionized species like 
Fe XXIV, Fe XXV or Fe XXVI, that are commonly observed in other dippers (for example EXO 0748-676: \citealt{Ponti2014}, 4U 1323-62 and 
4U 1624-490: \citealt{Boirin}). However, they have derived upper limits on the equivalent width of the 6.65 keV absorption line
corresponding to the Fe XXV 1s–2p transition, as 114, 50, and 73 eV in the persistent, shallow-dipping, and deep-dipping spectra, 
respectively.\\

We conservatively identify the edge feature at $\sim$2.8 keV in the \chandra spectra (within errors) as highly ionzed Sulphur 
species, S XVI, which has been observed in MR 2251-178 \citep{Gofford2011}, EXO 0748-676 \citep{Ponti2014}, GRS 1915+105 \citep{Ueda2009} 
and also in Cyg X-1 \citep{Feng2003}. It is likely that, as seen in EXO 0748-676, the highly ionized absorber on the 
accretion disc, could be responsible for the S XVI absorption edge as well.\\

Such high ionization Fe lines are tracers for equatorial disc winds \citep{Ponti2014}. The lines are suspected to be 
produced by absorption due to photo-ionized gas \citep{Ponti2015}. The Fe XXV K$\alpha$ (E = 6.697 keV) and Fe XXVI K$\alpha$ 
(E = 6.966 keV), as well as Fe XXV K$\beta$ (E = 7.880 keV) absorption lines have been observed in three other eclipsing 
sources AX J1745.6-2901 \citep{hyodo}, EXO 0748-676 \citep{Ponti2014} and IGR J17451-3022 \citep{Bozzo}. \citet{hyodo} 
suggested a disc-corona origin for these lines in the source AX J1745.6-2901, since they were observed at all orbital phases. 
Other non-eclipsing, high inclination dippers like 4U 1254-690, 4U 1624-490 and MXB 1659-298 also show these features, 
suggesting an equatorial origin for these photo-ionized absorbing gas \citep{Ponti2014}. In black hole binaries, these absorption 
features were found to have high outflow velocities and therefore were assumed to be equatorial winds 
(see \citealt{Ponti2012a, Ponti2014}). \\

In a recent review, \citet{TrigoBoirin2016} show that, from a total of 18 LMXBs exhibiting presence of photo-ionized plasma in the form 
of absorption lines of Fe XXV and Fe XXVI, 16 systems have highly ionized plasma (log($\xi$) $>$ 3). However, there are 4 systems 
that exhibit \textit{both} high (log($\xi$) $>$ 3) and low (log($\xi$) $<$ 3) photo-ionization and two systems 
(MAXI J1305-704 and GX 339-4) that show only low photo-ionization. Results obtained from the \suzaku observation in this work,
indicate that XTE J1710-281 could be one such system whose photo-ionized plasma has low ionization. Such photo-ionized 
plasma can be irradiation-induced, or even be the result of thermal or magnetic instabilities, etc.


\subsection{Nature of the absorber}

Various models have been proposed to explain the continuum emission in dippers. In order to particularly describe the 
comptonized component, a number of single and multi-component continuum powerlaw models have been previously explored.
One set of models (\citealt{Boirin,diaztrigo2006}) incorporate a neutral or ionized absorber partially or completely covering the power 
law as well as the blackbody emission during dips. These models explain the dip spectra in sources like 4U 1323-62, EXO 0748-676 and 4U 1254-690.
Another set of models (\citealt{Parmar2,bonnet,Oosterbroek,Younes}) incorporate two-components.
In these models, one fraction of the power law component is partially covered by an ionized or a neutral absorber and the 
remaining fraction is left uncovered. These models have been invoked to describe the dipping spectra in EXO 0748-676. Finally, 
the third set of models are the complex-continuum models (\citealt{church1998,church1999}). These models require two 
independent absorbing components, one for the extended comptonized power-law emission and another, for the point-like blackbody emission. 
Again, these complex models allow for neutral as well as ionized absorbers for the comptonized continuum component. Dippers like 
4U 1323-62 were modeled in this manner \citep{Boirin}. \\

In the case of XTE J1710-281, the spectral properties of the dips are well described by an ionized absorber that 
partially obscures the non-thermal emission and does not require us to invoke other complicated models. \\

Considering that the ionized absorber is located at the outer accretion disc, we can compute the ratio of thickness d, of a
slab of the absorber, with respect to the distance r, between the ionizing source and the absorber. For this, we consider 
the circularization radius to be the minimum possible value for the accretion disc radius, r, and compute it using equation (4.21) 
\citep{FKR} for a system with mass ratio, q$\sim$0.3. We obtain R$_{circ}$ $\sim$ 0.32R$_{\odot}$. 
Since $\xi$ = L/n$_{e}r^2$, and n$_{e} \sim$ ${n_{H}}^{zxipcf}$/d, we compute the ratio, \\

\begin{equation}
d/r = \frac{r\times\xi\times{n_{H}}^{zxipcf}}{L} \nonumber
\end{equation}
where, $\xi$ is the ionization parameter, L is the source luminosity, 
n$_e$ is the electron number density and ${n_{H}}^{zxipcf}$ is the ionized absorber column density. Assuming a distance of 15 kpc to 
the source \citep{Markwardt_2001} and a considering a flux of 7.9$\times$10$^{-11}$ ergs/s/cm$^{-2}$ (Table \ref{tab:avgspec}), we obtain a source luminosity of 
2$\times$10$^{36}$ ergs/s. This gives us a d/r ratio of 0.03 and 0.007 for the persistent and dipping intervals, respectively. \\

 Also, there is strong reason to believe that the absorber responsible for the X-ray 
intensity dips, also causes the Fe line absorptions as well, mainly because these lines are observed only in high inclination 
dipping sources. The relative line strengths in these different intensity epochs (persistent and dipping), 
can allow us to probe the absorber characteristics. For example, lines detected during the dips, correspond to electronic transitions 
from less ionized species than during the persistent \citep{Younes}. However we were unable to compare the line parameters in detail 
because of the weak line depths and statistical limitations of the dataset. \\

\section{Conclusions}

We have studied the eclipsing NS X-ray binary XTE J1710-281 using archival \chandra and \suzaku observations. 
Using \chandra folded orbital profiles, we detected an evolution of the dip morphology from orbit-to-orbit, indicating accretion disc 
structures, which seem to evolve in time-scales as short as a few hours. The intensity resolved \chandra and \suzaku continuum 
spectra are best described using a power-law that is partially absorbed by ionized material. 
We have detected signatures of accretion disc winds in the \suzaku spectra, in the form of highly ionized Fe absorption lines
at $\sim$6.60 keV and $\sim$7.01 keV and a broad Fe emission line complex at 0.72 keV. In the \chandra spectra, we detect the presence of an absorption edge at 2.8$\pm$0.2 keV, corresponding to the 
highly ionized S XVI species. Fe lines are detected only in \suzaku spectrum, but we cannot rule out their presence in 
the \chandra observation (as we mention) and in the \xmmn (as \citealt{Younes} mention). There is no clear correlation between 
the flux, shape of the continuum and presence/absence of lines in these observations. The shape of the continuum is consistent 
between the different observations: the case of the \xmmn observation \citep{Younes}, and also for the case of \chandra and 
\suzaku observations (present work). Future observations of this object will help understand the distribution of ionized matter 
on the accretion disc and disc structure evolution scenarios better.\\

\textit{Acknowledgements}: This work has made use of data obtained from the High Energy Astrophysics Science Archive 
Research Center (HEASARC), provided by the NASA Goddard Space Flight Center. We thank the referee for constructive comments and suggestions.

\bibliography{bibtex}{}
\bibliographystyle{mn2e}

\end{document}